\documentclass[prd,nofootinbib,floats,superscriptaddress,eqsecnum,tightenlines,preprintnumbers,11pt]{revtex4}

\usepackage{hyperref}
\usepackage{graphicx}
\usepackage{amsmath,amssymb,amsfonts,amsthm,latexsym}
\usepackage{marginnote}
\usepackage{xcolor}
\usepackage{physics}

\def\be{\begin{equation}}
\def\ee{\end{equation}}
\def\beq{\begin{equation}}
\def\eeq{\end{equation}}
\newcommand{\bea}{\begin{eqnarray}}
\newcommand{\eea}{\end{eqnarray}}
\def\bi{\begin{itemize}}
\def\ei{\end{itemize}}
\def\ba{\begin{array}}
\def\ea{\end{array}}
\def\bfig{\begin{figure}}
\def\efig{\end{figure}}

\newcommand\rs{\mathfrak{r}_{\! s}}
\newcommand\rst{{r}_{\! s}}
\newcommand\mass{r_{\rm m}}

\newcommand{\al}{\varpi}
\newcommand{\bet}{\varrho}

\newcommand{\bb}{f_1}
\newcommand{\gp}{{p_1}}
\newcommand{\rz}{r_-}
\newcommand{\rp}{r_+}

\newcommand{\gb}{\mathfrak{g}}
\newcommand{\Ab}{\mathfrak{A}}
\newcommand{\Bb}{\mathfrak{B}}
\newcommand{\Cb}{\mathfrak{C}}
\newcommand{\geff}{g} 
\newcommand{\Ae}{A}
\newcommand{\Be}{B}
\newcommand{\Ce}{C}

\newcommand{\lc}{\kappa}
\newcommand{\rss}{{\rho}}
\newcommand{\rstarmin}{r_{*{\rm min}}}

\begin{document}

\title{Axial perturbations of black holes in scalar-tensor gravity:\\
 near-horizon behaviour}

\author{Karim Noui}
\affiliation{Laboratoire de Physique des deux Infinis IJCLab, Universit\'e Paris Saclay, CNRS, France}
\affiliation{Universit\'e Paris Cit\'e, CNRS, Astroparticule et Cosmologie, F-75013 Paris, France}
\author{Hugo Roussille}
\affiliation{Univ Lyon, ENS de Lyon,  CNRS, Laboratoire de Physique, F-69342 Lyon, France}
\author{David Langlois}
\affiliation{Universit\'e Paris Cit\'e, CNRS, Astroparticule et Cosmologie, F-75013 Paris, France}

\date{\today}

\begin{abstract}
We consider axial (or odd-parity) perturbations of non-spinning hairy black holes (BH) in shift-symmetric DHOST (Degenerate Higher-Order Scalar-Tensor) theories, including terms quartic and cubic in second derivatives of the scalar field.
We give a new formulation of  the effective metric in which axial perturbations propagate as in general relativity.  We then introduce  a generic parametrization of the effective metric in the vicinity of the background BH horizon. Writing the dynamics of the perturbations in terms of a Schr\"odinger-like operator, we discuss in which cases the operator is (essentially) self-adjoint, thus leading to an unambiguous time evolution, according to the choice of  parameters characterizing the near-horizon effective metric. This is in particular useful to investigate the stability of the perturbations.
 We finally illustrate  our general analysis with two examples of BH solutions.
 \end{abstract}

\maketitle


\section{Introduction}

Future observations of binary black hole  mergers via GWs offer the tantalizing prospect of detecting the oscillations of the newly created back hole (BH) in the so-called post-merger ringdown phase. The main contribution to these oscillations can be decomposed into discrete modes  known as quasi-normal modes (QNMs) -- instead of normal modes because these modes decay as they are radiated away.  The measurement of QNM frequencies and decay rates provides a very powerful test of general relativity (GR) in the vicinity of a black hole.  Indeed, the frequencies and decay rates of QNMs (labelled by the integers $\ell$, $m$ and $n$)  depend only on the two parameters of a Kerr black hole solution, namely its mass and angular momentum (assuming the BH to be electrically neutral). The measurement of the frequency $\sigma$ and the decay rate $\tau$ of a single QNM, associated with the complex frequency $\omega=\sigma + i/\tau$ thus completely determines the two parameters and consistency can be checked by measuring any other QNM.  (Note that the astrophysical environment of the BH could potentially affect the QNMs spectrum and lead to an instability, see e.g. \cite{Jaramillo:2020tuu,Jaramillo:2021tmt}.)

With these observational prospects in mind, it is interesting to explore possible deviations  from the GR predictions within models of modified gravity. The theory of BH perturbations is already quite involved in general relativity (see e.g. \cite{Chandrasekhar:1985kt}). In the context of modified gravity, the analysis is even more daunting, due to different factors: paucity of exact  BH solutions, in particular of rotating BHs; presence of extra modes (e.g. scalar modes in scalar-tensor theories); modified dynamical equations.
In the present work, we avoid some of these difficulties by concentrating on nonrotating configurations and on purely tensorial modes. In the context of scalar-tensor theories, this means that we focus our attention on axial, or odd-parity, modes which are analogous to those of GR but now possess different dynamics. By contrast, we do not discuss here the polar, or even-parity, modes, which contain additional modes due to the presence of the scalar field perturbations.

In the present study, 
we work in the most general framework of scalar-tensor theories propagating a single scalar degree of freedom, known as DHOST (Degenerate Higher-Order Scalar-Tensor) theories~\cite{Langlois:2015cwa,Langlois:2015skt,BenAchour:2016cay,Crisostomi:2016czh,BenAchour:2016fzp} (see \cite{Langlois:2018dxi} and  \cite{Kobayashi:2019hrl} for reviews), which include previously studied families of scalar-tensor theories. General DHOST Lagrangians contain second derivatives of the scalar field and can lead to equations of motion of order higher than two, but imposing the degeneracy conditions ensures that the solutions of these equations of motion do not depend on more initial conditions than standard second-order equations of motion. In other words, these degeneracy conditions guarantee the absence of a ghost-like extra degree of freedom. 

Perturbations of BH in the context of DHOST theories have been explored in several works (see e.g. \cite{Kobayashi:2012kh,Cisterna:2015uya,Takahashi:2016dnv,Ganguly:2017ort, Takahashi:2019oxz,deRham:2019gha,Khoury:2020aya,Chatzifotis:2021pak,Takahashi:2021bml,Kase:2021mix,Tomikawa:2021pca,Nakashi:2022wdg,Minamitsuji:2022vbi,Minamitsuji:2022mlv}). One can also mention other works based on an EFT approach  \cite{Franciolini:2018uyq,Hui:2021cpm,Khoury:2022zor,Mukohyama:2022enj,Mukohyama:2022skk}.
In particular, in our previous works \cite{Langlois:2021xzq,Langlois:2021aji},  we have explored the equations of motion and the asymptotic behaviours of the solutions. We have also discussed the effective metric in which odd-parity modes propagate~\cite{Langlois:2022ulw}. The present work extends these results in two directions.

First, we consider  the effective metric for odd-parity modes in DHOST theories including cubic terms. Interestingly, we find that this metric can be partially written in a covariant form, even if the nice geometric interpretation of the effective metric as the result of a disformal transformation in quadratic DHOST is no longer valid when cubic terms are included\footnote{Note that generalised disformal transformations have recently been discussed in  \cite{Babichev:2021bim,Takahashi:2021ttd,Takahashi:2022mew,Ikeda:2023ntu}.}.

Second, assuming a generic power-law behaviour of the effective metric near the BH horizon, we discuss the self-adjointness  of the Schr\"odinger-like equation describing the odd-parity modes. Interestingly, a singular effective metric, corresponding to a naked singularity, does not necessarily lead to a loss of predictability for the associated Schr\"odinger-like equation.  This property was already pointed out for some naked singularity geometries in the context of general relativity \cite{Wald:1980jn,Horowitz:1995gi,Ishibashi:2003ap}  (see also \cite{Sadhu:2012ur} for an interesting example in scalar-tensor theories), and we use some of the formalism introduced in these previous works to study some generic behaviours of the effective metric.

This article is structured as follows. In the next section, we present the DHOST theories and introduce the effective metric for the axial perturbations of a static black hole solution. 
We then reformulate the dynamics of these perturbations in terms of a simple Schr\"odinger-like equation. Section \ref{section:asymptotics} introduces a general parametrization of the effective metric near the BH horizon and studies the self-adjointness of the Schr\"odinger-like, depending on the values of the parameters. The section concludes with a discussion on the stability of the axial perturbations. In Section \ref{section:Examples}, we illlustrate our general approach with two specific examples of BH solutions. 
We conclude in the final section. This article is completed with a few appendices, where the details of our calculations are presented.

\section{Effective metric for BH Axial Perturbations}

\subsection{Static and spherically symmetric geometries in DHOST Theories}
\label{subsec:ansatze}

As discussed in the introduction, we work within the framework  of DHOST theories, whose action, up to cubic order in second derivatives of the scalar field,   can be written in the form \cite{Langlois:2015cwa,BenAchour:2016fzp}
\begin{eqnarray}
	S[g_{\mu\nu},\phi] = \int \dd^4{x} \sqrt{-g} \Big(P(X,\phi) + Q(X,\phi) \square\phi + L^{(2)} + L^{(3)} \Big) \, ,
	\label{eq:generic-cubic-DHOST}
\end{eqnarray}
where $g_{\mu\nu}$ denotes the metric, $\phi$ the scalar field and $X\equiv \phi^\mu \phi_\mu$ its kinetic term, using  the short-hand notation $\phi_\mu=\nabla_\mu \phi$. The lagrangian densities $L^{(2)}$ and $L^{(3)}$ contain all the terms that are, respectively,  quadratic and cubic  in the second derivatives
$\phi_{\mu\nu} \equiv \nabla_\mu \nabla_\nu \phi$  and the associated curvature-dependent terms:
\begin{equation}
	L^{(2)} = F_2(X,\phi) R + \sum_{i=1}^5 {A}_i(X,\phi) L_i^{(2)} \,, \quad
	L^{(3)} = F_3(X,\phi) G_{\mu\nu} \phi^{\mu\nu} + \sum_{i=1}^{10} B_i(X,\phi) L_i^{(3)} \,,
	\label{DHOSTaction}
\end{equation}
where $R$ is the Ricci scalar and
$G_{\mu\nu}$  the Einstein tensor.
The five elementary quadratic Lagrangian  $L_i^{(2)}$  read
\begin{align}
	 & L^{(2)}_1 = \phi_{\mu \nu} \phi^{\mu \nu} \,, \qquad
	L^{(2)}_2 =(\Box \phi)^2 \,, \qquad
	L_3^{(2)} = (\Box \phi) \phi^{\mu} \phi_{\mu \nu} \phi^{\nu} \,,  \nonumber    \\
	 & L^{(2)}_4 =\phi^{\mu} \phi_{\mu \rho} \phi^{\rho \nu} \phi_{\nu} \,, \qquad
	L^{(2)}_5= (\phi^{\mu} \phi_{\mu \nu} \phi^{\nu})^2\, ,
\end{align}
while the ten  elementary cubic Lagrangian densities $L_i^{(3)}$  are given by \cite{BenAchour:2016fzp}
\begin{align}
	 & L^{(3)}_1=  (\Box \phi)^3  \,, \quad
	L^{(3)}_2 =  (\Box \phi)\, \phi_{\mu \nu} \phi^{\mu \nu} \,, \quad
	L^{(3)}_3= \phi_{\mu \nu}\phi^{\nu \rho} \phi^{\mu}_{\rho} \,,  \nonumber                                                              \\
	 & L^{(3)}_4= \left(\Box \phi\right)^2 \phi_{\mu} \phi^{\mu \nu} \phi_{\nu} \,, \quad
	L^{(3)}_5 =  \Box \phi\, \phi_{\mu}  \phi^{\mu \nu} \phi_{\nu \rho} \phi^{\rho} \,, \quad
	L^{(3)}_6 = \phi_{\mu \nu} \phi^{\mu \nu} \phi_{\rho} \phi^{\rho \sigma} \phi_{\sigma} \,, \nonumber                                   \\
	 & L^{(3)}_7 = \phi_{\mu} \phi^{\mu \nu} \phi_{\nu \rho} \phi^{\rho \sigma} \phi_{\sigma} \,, \quad
	L^{(3)}_8 = \phi_{\mu}  \phi^{\mu \nu} \phi_{\nu \rho} \phi^{\rho}\, \phi_{\sigma} \phi^{\sigma \lambda} \phi_{\lambda} \,,  \nonumber \\
	 & L^{(3)}_9 = \Box \phi \left(\phi_{\mu} \phi^{\mu \nu} \phi_{\nu}\right)^2  \,, \quad
	L^{(3)}_{10} = \left(\phi_{\mu} \phi^{\mu \nu} \phi_{\nu}\right)^3 \,.
\end{align}

The functions $P$ and $Q$ in \eqref{eq:generic-cubic-DHOST} can be chosen arbitrarily, while the other ones, $F_i$, $A_i$ and $B_i$, must satisfy  {\it degeneracy} conditions in order to guarantee the presence of a single scalar degree of freedom~\cite{Langlois:2015cwa}.  For theories up to cubic order, these degeneracy conditions were explicitly computed in \cite{BenAchour:2016fzp}, generalising the degeneracy conditions for quadratic DHOST theories established in \cite{Langlois:2015cwa}.

Note that, for axial modes, one does not need to take into account the degeneracy conditions since the scalar perturbations vanish by construction in the odd-parity sector. So our results would be unchanged for non degenerate scalar-tensor theories. This applies for example to U-DHOST theories~\cite{DeFelice:2018ewo}, theories that are degenerate in the so-called unitary gauge (where the scalar field is uniform) but not in other gauges,  or to the scordatura model~\cite{Motohashi:2019ymr}, in which the theory is a small controlled deformation of a DHOST theory so that the ghost-like extra degree of freedom is too massive to be excited in the regime of validity of the theory.

In the rest of this paper, we will  restrict our study to shift-symmetric theories, which entails that
all the  functions  in the action \eqref{eq:generic-cubic-DHOST}  depend on the kinetic density  $X$ only, and not explicitly on the scalar field $\phi$ itself.

\medskip

In the following, we consider static and spherically symmetric  solutions of these theories,  characterised by  a metric  $\gb_{\mu\nu}$ expressed as 
\begin{equation}
	\label{metric}
	\dd{s}^2 = \gb_{\mu\nu} \dd{x}^\mu \dd{x}^\nu = - \Ab(r) \dd{t}^2 + \frac{1}{\Bb(r)} \dd{r}^2 + \Cb(r)  ( \dd\theta^2 + \sin^2\theta \, \dd\varphi^2 ) \,,
\end{equation}
and a scalar field of the form
\begin{eqnarray}
	\label{generalscalar}
	\phi(t,r) = q t + \psi(r) \, ,
\end{eqnarray}
where $q$ is a constant. An explicit linear time dependence, i.e. $q\neq 0$,  initially proposed in  \cite{Babichev:2013cya} (see \cite{Mukohyama:2005rw} for an earlier work in a different context), is compatible with the assumption of staticity provided the theory is shift-symmetric (i.e. its Lagrangian depends only on the derivatives of $\phi$, not $\phi$ itself) and allows configurations with a time-like gradient for the scalar field.

Various exact BH solutions of this type have been obtained~\cite{Babichev:2017guv,BenAchour:2018dap,Motohashi:2019sen,Minamitsuji:2019shy,BenAchour:2020wiw,Minamitsuji:2019tet,Takahashi:2020hso} (see also the reviews \cite{Babichev:2016rlq,Kobayashi:2019hrl} on Horndeski theories and references therein). In the present work, we study generic properties of axial perturbations about
static and spherically symmetric solutions and, as such, we do not assume any specific solution.  We will nevertheless mention a few particular solutions  as illustrative examples in the final part. 

\subsection{Effective metric of axial perturbations}

As shown in several previous studies \cite{Tomikawa:2021pca,Takahashi:2021bml,Langlois:2022ulw},  axial modes  propagate in an effective metric $\geff_{\mu\nu}$ which in general is distinct from the background metric $\gb_{\mu\nu}$. Indeed,  in a scalar-tensor gravitational  theory,  axial gravitational waves are sensitive to 
 the background scalar field in addition to the background metric.  This affects their dynamics,  in contrast with other fields  that are  minimally coupled  to the metric $\gb_{\mu\nu}$, such as  an electromagnetic field with a standard action. As a consequence,  axial gravitational waves and, say, photons  effectively propagate in different geometries.

More specifically, we have shown in \cite{Langlois:2022ulw} that, for {\it axial} perturbations about  a static and spherically symmetric
background solution of the form \eqref{metric}-\eqref{generalscalar},   there exists a correspondence between their dynamics  in
any higher-order scalar-tensor theory of the form  \eqref{eq:generic-cubic-DHOST} (even non degenerate) and the usual   GR dynamics  but in an effective background metric of the form\footnote{Note that \eqref{effectivemetricform} has a priori no reason to be a {\it vacuum} solution of General Relativity, but  it can always be seen as a GR solution with appropriate, although artificial in general,   energy-momentum tensor.}
\begin{eqnarray}
	\label{effectivemetricform}
	\geff_{\mu\nu} \dd{x}^\mu \dd{x}^\nu  = - \Ae(r) \dd{t}_*^2 + \frac{1}{\Be(r)} \dd{r}^2 + \Ce(r)  ( \dd\theta^2 + \sin^2\theta \, \dd\varphi^2 ) \,,
\end{eqnarray}
with a new time coordinate $t_*$  defined by 
\begin{eqnarray}
	\label{defoftstar}
	t_* = t - \int \mathrm{d}r \,  \Psi(r) \,.
\end{eqnarray}
The effective metric components $\Ae$, $\Be$ and $\Ce$, as well as $\Psi$,  depend on the functions introduced in the  Lagrangian and on the background configuration \eqref{metric}-\eqref{generalscalar}. Their explicit expression for DHOST theories up to cubic order  was computed in \cite{Langlois:2022ulw} and is recalled in Appendix \ref{Section:effective}, where we also present  their extensions to non degenerate theories. 

Interestingly, after a long calculation summarised in the appendix, we find that this effective metric can be written in the compact and almost covariant form\footnote{In the general case,  many of the terms on the right-hand side involve the background metric $\gb_{\mu\nu}$ in a non-trivial way  and the relation \eqref{covarianteffective} defining
$\geff_{\mu\nu}$ in terms of $\gb_{\mu\nu}$ is presumably non invertible. It is nevertheless invertible in the particular case of quadratic theories, where the relation reduces to a disformal transformation. In this special case, the dynamics of the axial perturbations is equivalent  to that  of axial perturbations with respect to the effective metric in GR, as discussed in \cite{Langlois:2022ulw} and below.}
\begin{eqnarray}
	\label{covarianteffective}
	\geff_{\mu\nu} = \Lambda \left[ \Omega \, \gb_{\mu\nu} \, + \, D \, \phi_\mu \phi_\nu \, + \, S \, \phi_{\mu\nu} + \frac{T}{2} \left( \phi_\mu X_\nu + \phi_\nu X_\mu \right) \right] \, ,
\end{eqnarray}
where  $\Omega$, $D$, $S$ and $T$ can be simply expressed in terms of  the functions entering the action \eqref{DHOSTaction},
\begin{eqnarray}
	&&T = F_{3X} - \frac{3}{2} B_3 \, , \quad
	S = \frac{3}{2} X B_3 \, , \nonumber \\
	&& D = A_1 + (B_2 + \frac{2}{3} B_3) \Box \phi  + B_6\,  \phi^\alpha  \phi_{\alpha \beta} \phi^\beta \, , \label{expressionofDST}\\
	&&\Omega = F_2 - X D  -  T\,  \phi^\alpha  \phi_{\alpha \beta} \phi^\beta\,, \nonumber
\end{eqnarray}
evaluated on the background solution. Here, the $X$  subscript means a derivative with respect to $X$.
The global conformal factor $\Lambda$  takes the form
\begin{eqnarray}
	\Lambda \; = \; \Phi \sqrt{\frac{\Bb}{\Ab} \Gamma} \, ,
\end{eqnarray}
where $\Phi$ and $\Gamma$ were also introduced in \cite{Langlois:2022ulw} and are recalled in Appendix \ref{Section:effective}.

\medskip

As already stressed, the  expression \eqref{covarianteffective}-\eqref{expressionofDST} of the effective metric 
 is valid for any higher derivative scalar-tensor theories (up to the cubic order) without imposing the degeneracy conditions.   In the degenerate case, the term $\Box \phi$ disappears from \eqref{expressionofDST} because one of the degeneracy conditions imposes $2 B_2+ 3 B_3=0 $.
 
In the simpler case of quadratic theories (i.e. assuming that the cubic terms vanish: $F_3=B_2=B_3=B_6=0$), the terms $T$ and $S$ disappear and one recovers the expression of the effective metric obtained in \cite{Langlois:2022ulw} with
\begin{eqnarray}
	\Lambda = \sqrt{\frac{F_2}{ F_2 - X A_1}} \, , \quad  \Omega = F_2 - X A_1 \, , \quad D = A_1 \,.
\end{eqnarray}
As shown in \cite{Langlois:2022ulw}  the effective metric in this case  corresponds to the 
 disformal transformation of the background metric so that the new action, expressed in terms of $\geff_{\mu\nu}$, is characterised by the  functions 
 \begin{eqnarray}
	\tilde F_2 \, = \, 1 \, , \qquad
	\tilde A_1 \, = \, 0 \, .
\end{eqnarray}
As a consequence,  the dynamics of axial perturbations in the frame of the effective
metric is strictly equivalent to the dynamics of these perturbations in General Relativity. In other words, the effective metric can be interpreted as 
an ``Einstein frame" for axial perturbations. 
We stress that this does not mean that the gravitational theory is equivalent to General Relativity. First, because the matter fields, or other fields, are minimally coupled to the metric $\gb_{\mu\nu}$ and therefore nonminimally coupled to the effective metric. Second, because the other quadratic Lagrangian functions $\tilde A_i$ do not necessarily vanish, in which case  the dynamics of the polar perturbations in the effective metric frame differs from the GR one.

\subsection{Schr\"odinger-like equation for axial perturbations}
As recalled in the previous sections, the dynamics of axial perturbations in higher-order scalar-tensor theories (not necessarily degenerate) corresponds to that  of General Relativity with the effective metric \eqref{effectivemetricform} instead of  the background metric.

The GR dynamics of linear perturbations $h_{\mu\nu}$ about a reference metric ${g}_{\mu\nu}$ is given by the linearised Einstein equations,
\bea
{\cal E}_{\mu\nu} &\equiv   &{\Box} h_{\mu\nu} + \nabla_\mu \nabla_\nu h + (\nabla_\alpha \nabla_\beta h^{\alpha\beta}- \Box h) g_{\mu\nu} + 2 \nabla_{(\mu} \nabla_\alpha h^\alpha_{\nu)}  -6 \nabla_\alpha \nabla_{(\mu} h_{\nu)}^\alpha   \nonumber \\
&+&  R_{\mu\nu} h - R h_{\mu\nu} + \frac{1}{2} R g_{\mu\nu} h + R^{\alpha \beta}  g_{\mu\nu}  h_{\alpha \beta}
+ 8 R_{\alpha(\mu} h^\alpha_{\nu)}=0 \, ,
\label{eom}
\eea
where $h\equiv g^{\mu\nu}h_{\mu\nu}$ is the trace of $h_{\mu\nu}$ and we use the standard notation $A_{(\mu\nu)} \equiv (A_{\mu\nu} + A_{\nu\mu})/2$ for any 2-index tensor $A_{\mu\nu}$. In the above equation, indices are lowered or raised with the reference metric $g_{\mu\nu}$; the covariant derivatives are associated with  $g_{\mu\nu}$ as well. In the following, the nonperturbed metric will correspond to our effective metric.

From the above linearised Einstein's equations, one can derive  a Schr\"odinger-like equation for axial perturbations, as we now recall briefly. Details can be found in \cite{Regge:1957td} or in the more recent articles \cite{Kobayashi:2012kh,Langlois:2021aji,Langlois:2022ulw}. In the Regge-Wheeler gauge \cite{Regge:1957td}, axial perturbations
are described by the following non-vanishing components of the perturbations:
\begin{align}
	 & h_{t\theta} = \frac{1}{\sin\theta}  \sum_{\ell, m} h_0^{\ell m}(t,r) \partial_{\varphi} {Y_{\ell m}}(\theta,\varphi), \qquad
	h_{t\varphi} = - \sin\theta  \sum_{\ell, m} h_0^{\ell m}(t,r) \partial_{\theta} {Y_{\ell m}}(\theta,\varphi), \nonumber         \\
	 & h_{r\theta} =  \frac{1}{\sin\theta}  \sum_{\ell, m} h_1^{\ell m}(t,r)\partial_{\varphi}{Y_{\ell m}}(\theta,\varphi), \qquad
	h_{r\varphi} = - \sin\theta \sum_{\ell, m} h_1^{\ell m}(t,r)  \partial_\theta {Y_{\ell m}}(\theta,\varphi), \label{eq:odd-perttext}
\end{align}
using an expansion in spherical harmonics ${Y_{\ell m}}(\theta,\varphi)$, reflecting the spherical symmetry of the background. In the following, since perturbations with different values of $\ell$ and $m$ do not couple at  linear level, we drop the indices $\ell$ and $m$ to shorten the equations. 
Moreover,  we consider only $\ell \geq 2$ since axial perturbations contain no monopole ($\ell=0$) nor dipole ($\ell=1$) contributions.  

For $\ell \geq 2$, one can show that  only  three out of the ten equations  \eqref{eom} are non trivial and that only two of these three equations are independent, providing two equations for the  two functions $h_0$ and $h_1$.
Combining these two equations, and working in frequency space, so that 
any function $f(t_*,r)$ 
 is replaced by
\begin{eqnarray}
	f(t_*,r) \; = \; e^{-i \omega t_*} \, f(r) \, ,
\end{eqnarray}
we obtain the well-known Schr\"odinger-like equation\footnote{In addition to the Schr\"odinger-like equation, we obviously have a second equation whose expression is not needed here. However, these two equations are necessary to solve completely the perturbations equations, and then to find $h_0$ and $h_1$.}
\begin{eqnarray}
	\label{Schrooperator}
	-\frac{\partial^2 \chi}{\partial r_*^2} \, + \, V \, \chi = \omega^2 \, \chi \,.
\end{eqnarray}
The function $\chi(r)$ corresponds to the following linear combination of $h_0$ and $h_1$,
\begin{eqnarray}
	\label{defofchi}
	\chi  \; = \; \left(\frac{\cal F}{\Ce}\right)^{1/2} \left(\frac{\Be} {\Ae\Gamma}\right)^{1/4}\,  \frac{h_1 + \Psi h_0}{\omega N}   \, ,
\end{eqnarray}
where the definitions of $\cal F$, $\Gamma$ and $\Psi$  are recalled in Appendix \ref{Section:effective}.
The tortoise coordinate $r_*(r)$ 
for the effective metric  is defined as\footnote{We stress that $r_*$ does not correspond to the tortoise coordinate of the background metric.}
\begin{eqnarray}
	\label{tortoise}
	r_*=\int \frac{\dd{r}}{\sqrt{\Ae(r) \Be(r)}}\,.
\end{eqnarray}
Finally, the potential $V(r)$ can be written in the simple form
\begin{eqnarray}
\label{decomposition_potential}
	V =  S^2 - \partial_* S + V_\lambda \, , \qquad V_\lambda = 2\lambda \frac{\Ae}{\Ce} \,,\qquad
	S \equiv \frac{1}{2}  \frac{\partial_* \Ce}{\Ce}\,,
\end{eqnarray}
with
\begin{equation}
	\lambda \equiv  \frac{\ell (\ell +1)}{2} - 1 \,,
	\label{lambda}
\end{equation}
and where $\partial_*$ denotes a derivative with respect to $r_*$.

\section{Axial perturbations: asymptotics and stability}
\label{section:asymptotics}

\subsection{Asymptotic behaviour of the effective metric at the singularity}
As our starting point we consider  a  black hole background metric   with (at least) one event horizon located at $r=\rs $ in the static coordinates system
used in \eqref{metric}. The  corresponding effective metric $\geff_{\mu\nu}$ might not be that of a black hole: a priori, it can describe any static and spherically symmetric geometry, such as, for instance, a naked singularity, a regular space-time or a black hole whose horizon differs from the background one.  Different examples of such effective metrics have been considered in \cite{Langlois:2022ulw} and some of them will be discussed later on.

Here, we assume the effective metric to be singular (i.e. one its components either vanishes or diverges) for some value of the radius $r=\rst$, which can be different from $\rs$ or not. The singularity at  $r=\rst$ can be  a naked singularity or  a coordinate singularity.
Furthermore, we assume the metric to be well-defined and well-behaved in the domain $r>\rst$, by which we mean that all coefficients $\Ae(r)$, $\Be(r)$ and $\Ce(r)$ remain strictly positive for $r>\rst$,  and give an asymptotically flat metric.


To describe the  behaviour of the metric near the singularity, it is convenient to introduce the dimensionless coordinate $\varepsilon$ defined by
\begin{eqnarray}
	\varepsilon = \frac{r}{\rst} - 1 \,.
\end{eqnarray}
In the following, we will assume that  the effective metric coefficients  behave as  power laws of  $\varepsilon$  in the vicinity of the singularity, so that  their leading order terms  is of the form
\begin{eqnarray}
	\Ae(r) \, = \,  a \, \varepsilon^{\alpha} \left( 1+ o(1)\right)\, , \quad
	\Be(r) \, = \, b \, \varepsilon^{\beta} \left( 1+ o(1)\right) \, , \quad
	\Ce(r) \, = \, c \, \varepsilon^{\gamma} \left( 1+ o(1)\right) \,  \ {\rm when }\  \varepsilon \to 0\,,
	\label{eq:exponents}
\end{eqnarray}
where $\alpha$, $\beta$ and $\gamma$ are real constants, and $a$, $b$ and $c$ are  non-negative  numbers.

Rewriting the effective metric in terms of the tortoise coordinate \eqref{tortoise},
\begin{eqnarray}
	\label{metrictortoise}
	\dd s^2 \; = \; \Ae \left[ -\dd t_*^2 + \dd r_*^2 + \frac{\Ce}{\Ae} \left(\dd \theta^2 + \sin^2 \theta \, \dd  \varphi^2 \right) \right] \,,
\end{eqnarray}
the  near-singularity behaviour of the effective metric is described by 
\begin{eqnarray}
	\label{param2}
	\dd  s^2 \; \simeq \; a \, \varepsilon^{\alpha} \left[ - \dd t_*^2 + \dd r_*^2 + \lc^2 \, \varepsilon^\xi \left(\dd \theta^2 + \sin^2 \theta \, \dd  \varphi^2 \right) \right] \, , \qquad \dd r_* = \rss \, \varepsilon^{\sigma -1} \, \dd\varepsilon \, ,
\end{eqnarray}
where we have introduced the new parameters
\begin{eqnarray}
\label{asymptotic_parameters}
	\lc^2 \, \equiv \, \frac{c}{a} \, , \qquad \rss \,\equiv \,\frac{\rst}{\sqrt{a  b}} \, , \qquad
	\sigma \, \equiv \, 1 - \frac{\alpha+\beta}{2} \, , \qquad
	\xi \, \equiv \, \gamma - \alpha \, .
\end{eqnarray}

\medskip
In order to determine the nature of the singularity of the effective metric, it is useful to express $r_*$ in terms of $\varepsilon$  near the singularity. This gives
\begin{eqnarray}
\label{behaviour_tortoise}
	{\text{if}} \,\,\, \sigma = 0 \,, \,\,\ r_* \simeq  \, \rss \ln \varepsilon \, , \qquad
	{\text{if}} \,\,\, \sigma \neq 0 \,, \,\,\ \sigma \, r_* \simeq {\rss }  \, \varepsilon^\sigma \, .
\end{eqnarray}
When $\sigma \leq 0$, the singularity $r=\rst$ corresponds to the limit  $r_* \rightarrow -\infty$ and the
domain of the Schr\"odinger-like equation \eqref{Schrooperator} is  therefore the full real line $\mathbb R$. As a consequence, following the arguments of \cite{Horowitz:1995gi}, the singularity is null  and the space-time is globally hyperbolic.  By contrast, when $\sigma >0$, the singularity is located at  $r_*=0$.  This gives 
a time-like singularity and the domain of  the Schr\"odinger-like equation is reduced to the half-line $\mathbb R_*^+$.

As argued  in \cite{Horowitz:1995gi}, one expects the Schr\"odinger operator in \eqref{Schrooperator} to be essentially self-adjoint
when the singularity is null, which means that the evolution of the perturbation is well-defined and does not need any extra boundary conditions at the singularity. However, this might not be the case when the singularity is time-like and a careful study of the asymptotic behaviour of the potential at the singularity is needed to conclude about the self-adjointness.

\subsection{ Asymptotic of the potential and self-adjointness of the Schr\"odinger operator}

The detailed study of the asymptotic behaviour of the potential at the singularity is given in Appendix  \ref{App:asymp}. This is crucial  to  conclude about the self-adjoint properties of the Schr\"odinger operator \eqref{Schrooperator} and the positivity of its spectrum, which in turn, will enable us to conclude on the stability of axial perturbations  (see e.g. \cite{Lewin2022} for a general study of the Schr\"odinger operator).

We summarise below the main results of  Appendix  \ref{App:asymp}. We treat separately the cases $\sigma=0$, $\sigma<0$ and $\sigma >0$ and we assume that $\gamma \neq 0$ and $\gamma + 2\sigma \neq 0$. The special cases $\gamma = 0$  and $\gamma + 2\sigma=0$  are studied in the appendix.

\subsubsection{Case $\sigma=0$  (and $\gamma\neq 0$)}
According to \eqref{behaviour_tortoise}, the singularity is null, located at $r_* \rightarrow - \infty$.  Let us examine in turn the cases $\xi > 0$ and  $\xi \leq 0$.

\medskip
For $\xi>0$,  the potential  behaves near the singularity as
\begin{eqnarray}
	V \; \simeq \;   V_{s}^+ \, \exp (-   \xi \frac{r_*}{\rss}) \, ,  \qquad V_{s}^+= 2\frac{\lambda}{\lc^2}\qquad \quad (\ r_* \rightarrow - \infty, \  \xi > 0) 
\end{eqnarray}
so that  $\lim_{r_* \rightarrow - \infty} V(r_*) = + \infty$.
Using the results of  \cite{berezin1991one} on the asymptotic behaviour of the solutions of one-dimensional  Schr\"odinger
equations, summarized in Appendix \ref{App:SA}, we can find a pair of independent solutions $\chi_\pm$ which
behave at the singularity as follows,
\begin{eqnarray}
	\chi_\pm(r_*) \simeq  \, \exp \left[ \frac{\xi r_*}{4 \rss }\pm  \frac{2 V_s^{+1/2} \rss}{\xi} \exp \left( - \xi \frac{r_*}{2 \rss} \right)\right] \quad (r_* \rightarrow - \infty,\ \xi > 0) \;.
\end{eqnarray}
This leading behaviour does not depend on the frequency $\omega$, and one immediately sees  that $\chi_-$
is an element of $L^2(\mathbb R)$ whereas $\chi_+$ is not. As a consequence, it is not necessary to add any boundary
condition at the singularity which means that the Schr\"odinger operator is essentially self-adjoint.

\medskip
Let us now turn to the case $\xi \leq 0$. In the limit $r_* \rightarrow - \infty$, the potential is constant:
\begin{eqnarray}
	V \; \simeq \;  2\frac{\lambda}{\lc^2} \, \delta_\xi+ \frac{\gamma^2}{4 \rss^2} \equiv V_{s}^- 
\end{eqnarray}
with $\delta_\xi=1$ if $\xi=0$ and $\delta_\xi=0$  otherwise.  In this case, one can find  two solutions, still denoted $\chi_\pm$, behaving as
\begin{eqnarray}
	\chi_\pm(r_*) \simeq  \exp \left[ \pm \sqrt{V_s^- - \omega^2} \, r_* \right]  \quad (r_* \rightarrow - \infty, \ \xi \leq 0) \; ,
\end{eqnarray}
where  the square root is imaginary if $V_s-\omega^2<0$. To prove that the corresponding Schr\"odinger operator is self-adjoint, it is sufficient\footnote{More details  on this theorem can be found in \cite{berezin1991one}   for instance. Some illustrative examples, which are furthermore physically relevant, are studied in \cite{Horowitz:1995gi}.} to show that, when
$\omega = \pm i$, one of two solutions $\chi_\pm$ is not in $L^2(\mathbb R)$ which is clearly the case (since $V_s^- +1>0$).

\medskip
For $\sigma=0$, we thus conclude that the Schr\"odinger operator is essentially self-adjoint. This is consistent with the fact that the space time
is globally hyperbolic in this case.

\subsubsection{The case $\sigma<0$ (and $\gamma\neq 0, -2\sigma$)}
According to \eqref{behaviour_tortoise},  the singularity is still null and located at $r_* \rightarrow - \infty$. As found in Appendix \ref{App:asymp}, the behaviour of  the potential near the singularity is given by
\begin{eqnarray}
	\label{asympsigma}
	V \, \simeq \, V_{*s} \, \vert r_* \vert^{\nu} \quad \text{with} \quad  V_{*s} = V_s \left( \frac{ \vert \sigma \vert}{\rss}\right)^{\nu} \, ,
\end{eqnarray}
where 
\begin{eqnarray}
	&&\text{if} \; \xi > 2 \sigma \; :  \nu = - \frac{\xi}{\sigma}, \quad  V_{s} =  2 \frac{\lambda}{\lc^2}\, , \label{zeta1}\\
	&&\text{if} \; \xi \leq 2\sigma \; :  \nu =  -2  , \quad  V_{s} =  2\frac{\lambda}{\lc^2} \, \delta_{\xi-2\sigma}+ \frac{\gamma(\gamma + 2\sigma)}{4 \rss^2}  \, .\label{zeta2}
\end{eqnarray}

The asymptotic behaviour of the potential when $r_* \rightarrow - \infty$ depends on the sign of $\nu$. 
In each case, one can find two solutions $\chi_\pm$ whose asymptotic behaviour at the singularity can be obtained explicitly:
\begin{eqnarray}
	&& \nu < 0 \, , \quad \lim_{r_* \rightarrow - \infty} V(r_*) = 0 \, , \qquad \quad \chi_\pm(r_*) \simeq  \exp \left[ \pm i \, \omega \, r_* \right] \, ,\\
	&& \nu = 0 \, , \quad \lim_{r_* \rightarrow - \infty} V(r_*) = V_{*s} \, , \qquad    \;\; \chi_\pm(r_*) \simeq  \exp \left[ \pm  \sqrt{V_{*s}- \omega^2 } \, r_*\right]  \, ,\\
	&& \nu > 0 \, , \quad \lim_{r_* \rightarrow - \infty} V(r_*) = + \infty \, ,\qquad  \chi_\pm(r_*) \simeq  \frac{1}{\vert r_*\vert^{\nu/4}}  \, \exp \left[ \frac{ \pm 2 }{\nu +2} \sqrt{V_{*s}} \vert r_*\vert ^{1+\nu/2}\right] \,.
\end{eqnarray}
We see that, when $\omega = \pm i$, one of the two solutions does not belong to $L^2(\mathbb R)$, which is
a sufficient condition for the operator to be self-adjoint. Again, this is  consistent with the fact that the effective spacetime
is globally hyperbolic.

\subsubsection{The case $\sigma>0$ (and $\gamma\neq 0, -2\sigma$)}
According to \eqref{behaviour_tortoise},  the singularity is now time-like, located at $r_* =0$. 
The asymptotic behaviour of the potential can be written in the form \eqref{asympsigma}, but now with 
\begin{eqnarray}
	\label{sigmapositive}
	&&\text{if} \; \xi > 2 \sigma \; :  \nu < -2 \; \; \text{and} \;\; V_{s} >0 \, ; 
	\\
	&&\text{if} \; \xi \leq 2\sigma \; :  \nu =  -2 \,,\quad  V_{s} =  2\frac{\lambda}{\lc^2} \, \delta_{\xi-2\sigma}+ \frac{\gamma(\gamma + 2\sigma)}{4 \rss^2}  \, .\label{sigmaneg3}
\end{eqnarray}

For $\xi > 2 \sigma$,  following the usual method, one can  find two independent solutions $\chi_\pm$, with the following 
asymptotic behaviour near the singularity:
\begin{eqnarray}
	\chi_\pm(r_*) \simeq \frac{1}{\vert r_*\vert ^{\nu/4}} \, \exp \left[ \frac{\pm 2}{\nu + 2} \vert r_*\vert^{1+\nu/2}\right] \,,\qquad r_*\to 0\,,
\end{eqnarray}
which are formally the same expressions as those in \eqref{sigmaneg3}.
Since $1+\nu/2 < 0$ in this case, one concludes that  $\chi_-$ is an element of
$L^2(\mathbb R)$ whereas $\chi_+$ is exponentially divergent at the origin.

Similarly, for $\xi \leq 2\sigma$ one can easily  find two independent solutions, which behave asymptotically as 
\begin{eqnarray}
	\chi_\pm(r_*) \simeq  r_*^{ n_\pm} \, , \qquad n_\pm \equiv \frac{1 \pm \sqrt{1+4 V_{*s}} }{2} \,.
\end{eqnarray}
Combining \eqref{asympsigma} with \eqref{sigmaneg3} and using \eqref{asymptotic_parameters}, the constant $V_{*s}$ is explicitly given by
\begin{eqnarray}
	V_{*s} = 2 \frac{\lambda}{\sigma^2 } \frac{\rst^2}{b c} \delta_{\xi-2\sigma} +    \frac{\gamma(\gamma + 2\sigma)}{4 \sigma^2}  \, .
	\label{eq:def-Vstars}
\end{eqnarray}
When $V_{*s} \geq 3/4$, the solution $\chi_-$ is not square integrable near $r_*=0$ whereas $\chi_+$ is, so  the Schr\"odinger operator is essentially self-adjoint. When $V_{*s}<3/4$, both solutions are integrable near the singularity. This implies that  the operator is not essentially self-adjoint and  boundary conditions at the singularity are required in order  to know how the perturbation evolves from a given initial condition.  In this case, the singularity is time-like and the space-time is not globally hyperbolic.

\medskip
In summary, we have found that  the Schr\"odinger operator is essentially self-adjoint, except when
\begin{eqnarray}
	\sigma >0 \, , \quad \xi \leq 2 \sigma \, , \quad V_{*s} < \frac{3}{4} \,.
\end{eqnarray}
The last inequality can be reformulated as follows:
\begin{eqnarray}
	&&\text{if} \, \; \xi < 2 \sigma \, , \;\;\; -3 \sigma < \gamma < \sigma \, ; \qquad
	\\
	&&\text{if} \, \;  \xi = 2 \sigma \, , \;\;\; 
	8 \lambda \frac{\rst^2}{b c}< (3\sigma-\gamma)(\sigma+\gamma)\,.
\end{eqnarray}

\subsection{Stability of axial perturbations}
\label{subsec:stability}
We now discuss the stability of the axial perturbations, using arguments similar to those given in \cite{Kodama:2003kk}  (see also \cite{Ganguly:2017ort, Kimura:2018whv, Takahashi:2019oxz}). Stability can be shown by using the property  that the Schr\"odinger-like operator is a positive self-adjoint operator in the space of square integrable functions, which implies that there is no 
normalisable unstable mode.

One way to show stability is to show that $\omega^2>0$ for any square integrable function $\chi$ in the domain of the Schrödinger-like equation~\eqref{Schrooperator}.

\medskip
Multiplying  the Schr\"odinger equation \eqref{Schrooperator} by $\overline{\chi}$, one obtains
\begin{eqnarray}
	\label{SchroIntegrated}
I_\omega\equiv	\omega^2 \, \int_{\rstarmin}^{+\infty} \dd r_* \, \vert \chi \vert^2 = \int_{\rstarmin}^{+\infty} \dd r_* \left[ -\overline{\chi}\, \partial_*^2\chi+ (S^2-\partial_* S+V_\lambda) \vert \chi \vert^2 \right]\,,
\end{eqnarray}
where $\partial_*$ denotes a derivative with respect to $r_*$ and where we have used the decomposition \eqref{decomposition_potential}  of the potential.
By introducing  the derivative operator $D$ defined by
\begin{eqnarray}
	D \chi \equiv \partial_* \chi + S \, \chi \, ,
\end{eqnarray}
we can rewrite the above integral in the convenient form 
\beq
I_\omega= \int_{\rstarmin}^{+\infty} \dd r_* \left( \vert D \chi \vert^2 + V_\lambda \vert \chi \vert^2 \right) - [\overline{\chi} \partial_* \chi + S \vert \chi \vert^2]_{\rstarmin}^{+\infty} \,.
\eeq
If  the right-hand side can be shown to be positive then this implies  $\omega^2>0$ and therefore the stability of the mode.
Let us therefore analyse the boundary terms.

First of all, since the effective metric is supposed to be asymptotically flat, the boundary term for $r_* \rightarrow \infty$ vanishes and the only remaining 
  boundary term is
  \begin{eqnarray}
	\label{boundS}
	\lim_{r_* \rightarrow \rstarmin} \left( \overline{\chi} \partial_* \chi + S \vert \chi \vert^2 \right) \, .
\end{eqnarray}

When $\sigma \leq 0$, the singularity is sent to infinity, i.e. $\rstarmin=-\infty$. Using the explicit expression of $S$,
\beq
S = \frac{1}{2}  \frac{\partial_* \Ce}{\Ce}=\frac{1}{2} \sqrt{\Ae\Be}\,  \frac{\Ce'}{\Ce}\,,
\eeq
and substituting the behaviour of the effective metric functions near the singularity, one finds the following behaviours for  $S$, depending on the value of $\gamma$:
\begin{eqnarray}
	\text{if} \, \gamma \neq 0 \, , \;\; S \simeq \frac{\gamma}{2 \rss} \varepsilon^{-\sigma} \, , \qquad
	\text{if} \, \gamma = 0 \, , \;\; S \simeq \frac{c_1}{2 \rss} \varepsilon^{1-\sigma}  \, ,
\end{eqnarray}
where $c_1$ is introduced in \eqref{a1b1c1}. Hence, $S$ is bounded (it even goes to zero when $\sigma <0$) and the boundary term \eqref{boundS} vanishes, which implies
the stability of axial perturbations.

\medskip

When $\sigma >0$, the situation is different because the domain reduces to the real half-line, with the singularity located at $\rss=0$.  In this case, the function $S$ diverges at the singularity, according to
\begin{eqnarray}
	S \simeq \frac{\gamma}{2\sigma r_*} \,,
\end{eqnarray}
where $\gamma$ is supposed not to vanish. To go further, we need to distinguish between the cases $\xi > 2 \sigma$ and $\xi \leq 2 \sigma$ \eqref{sigmapositive}. In the former case (which corresponds to an essentially self-adjoint Schr\"odinger operator), we showed that the solution of the Schr\"odinger equation behaves as
\begin{eqnarray}
	\chi(r_*) \sim \frac{1}{r_*^{\nu/4}} \, \exp \left[ \frac{2}{\nu + 2} r_*^{1+\nu/2}\right] \, ,
\end{eqnarray}
where $\nu+2<0$, and then we restrict the functions in \eqref{SchroIntegrated} to behave exactly as the solutions at the singularity. As a consequence, the boundary term   in \eqref{SchroIntegrated}  vanishes and axial perturbations are stable.

In the latter case ($\xi \leq 2 \sigma$),  we need to treat separately the cases where the Schr\"odinger operator is essentially self-adjoint and where it is not. When it is essentially self-adjoint, i.e. $V_{*s} \geq 3/4$, the asymptotic behaviour of the solution is given by a power law
\begin{eqnarray}
	\chi(r_*) \sim  \, r_*^n \, , \qquad n = \frac{1 + \sqrt{1+4 V_{*s}}}{2} \, ,
\end{eqnarray}
where $n \geq 3/2$. Therefore, the boundary term vanishes and axial perturbations are stable.

When $V_{*s} <3/4$, the situation is different as the Schr\"odinger operator is not essentially self-adjoint and  boundary conditions at the singularity need to be prescribed.  Indeed, any solution behaves as follows at the singularity,
\begin{eqnarray}
	\chi(r_*) \, \simeq \,\alpha_+ r_*^{n_+} + \alpha_- r_*^{n_-} \, , \qquad n_\pm = \frac{1 \pm \sqrt{1+4 V_{*s}} }{2} \, ,
\end{eqnarray}
where $\alpha_\pm$ are constants. We immediately show that
\begin{eqnarray}
	\frac{1}{2} \leq \Re(n_+) < \frac{3}{2} \, , \qquad
	-\frac{1}{2} \leq \Re(n_-) < \frac{1}{2} \, ,
\end{eqnarray}
which confirms that both solutions are square integrable at the singularity. If we choose the boundary condition such that
$\alpha_+=0$, then the boundary term diverges at the singularity and then we cannot conclude that axial perturbations are stable
using the properties of the  integral \eqref{SchroIntegrated}. On the contrary, if we choose $\alpha_-=0$, the boundary term does not
diverge anymore and vanishes when $V_{*s} \neq 0$ which may be the sign of stability.

\section{Examples}
\label{section:Examples}

Let us now turn to the application of the above analysis  to specific black hole solutions in DHOST theories. We will consider two different solutions whose perturbations were studied in \cite{Langlois:2021aji, Langlois:2022eta}. These solutions span several subcases of the general theory and solution presented in Sec.~\ref{subsec:ansatze}: there are solutions with $q = 0$ and $q \neq 0$ and solutions of both quadratic and cubic DHOST theories.

\subsubsection{BCL solution}

The BCL solution, proposed in \cite{Babichev:2017guv} and named  here after its authors,  is a DHOST black hole solution whose metric is analogous to  Reissner-Nordström but with an imaginary charge. It constitutes an instructive toy model for the study of black holes in DHOST since it is different from the Schwarzschild solution but show comparable dynamics. This solution corresponds to the specific choice of DHOST functions
\begin{equation}
	F_2 = f_0 + f_1 \sqrt{X} \,,\quad P = - p_1 X \,,\quad A_1 = -A_2 = 2F_{2X} \,,
\end{equation}
while all other DHOST functions are zero. The metric is given by~\eqref{metric} with
\begin{align}
	&\Ab(r) = \Bb(r) = \qty(1 - \frac{r_+}{r}) \qty(1 + \frac{r_-}{r}) \,,\qquad \Cb(r) = r^2\,,\nonumber\\
	&\phi(r)=\psi(r)=\pm \frac{\bb}{\gp\sqrt{\rp\rz}}\arctan\left[\frac{\mass r + 2\rp\rz}{2\sqrt{ \rp\rz}\sqrt{(r-\rp)(r+\rz)}}\right]\, + {\text{cst}} \, .
\end{align}
where the (positive) quantities $r_+$ and $r_-$ are defined by
\begin{equation}
	r_+ r_- = \frac{f_1^2}{2f_0p_1} \,,\quad r_+ - r_- = 2 m \qq{and} r_+ > r_- > 0 \,,
\end{equation}
$m$ corresponding to the ADM mass. The metric has only one horizon located at $r = r_+$. The global sign of $\phi(r)$ and the constant are physically irrelevant \cite{Babichev:2017guv}. 

Axial perturbations of the BCL black hole and their effective metric were studied in \cite{Langlois:2021aji, Langlois:2022ulw}. The functions $\Ae$, $\Be$ and $\Ce$ are given by
\begin{align}
	\Ae(r) = f_0 \sqrt{1 + \frac{2r_+r_-}{r^2}} \, \Ab(r) \,,\quad \Be(r) = \frac{\Ab(r)}{f_0\qty(1 + \frac{2r_+r_-}{r^2})^{3/2}}  \,,\quad \Ce(r) = f_0 \sqrt{1 + \frac{2r_+r_-}{r^2}} r^2 \,,
\end{align} 
and the potential $V$ is
\begin{align}
	&V(r) = \Ab(r) \qty[ \frac{2\lambda}{r^2} + \frac{\sum_{n=0}^6 p_n r^n}{r^2(r^2 + 2 r_+ r_-)^3}] \,,\nonumber\\
	&p_0 = -5 r_+^3 r_-^3 \,,\quad p_1 = -3 r_+^2 r_-^2 (r_+ - r_-) \,,\quad p_2 = -6 r_+^2 r_-^2 \,,\nonumber\\
	&p_3 = - 4 r_+ r_- (r_+ - r_-) \,,\quad p_4 = -3 r_+ r_- \,,\quad p_5 = -3 (r_+-r_-) \,,\quad p_6 = 2 \,.
	\label{eq:potential-BCL}
\end{align}
Note that this expression differs from the potential given in \cite{Langlois:2021aji} because the radial coordinates used here and in that previous work are different.

We can study the asymptotic behaviour of perturbations of this solution using the framework developped in the present paper. In order to do this, we compute the asymptotics of the effective metric at the black hole horizon $r = r_+$. The quantities defined in~\eqref{eq:exponents} are
\begin{align}
	&\alpha = 1 \,,\quad \beta = 1 \,,\quad \gamma = 0\,,\quad \varepsilon = \frac{r}{r_+} - 1 \,,\nonumber\\
	&a = f_0 \sqrt{1 + \frac{2r_-}{r_+}} \frac{r_++r_-}{r_+} \,,\quad b = \frac{1}{f_0 \qty(1 + \frac{2r_-}{r_+})^{3/2}}\frac{r_++r_-}{r_+} \,,\quad c = f_0 r_+^2 \sqrt{1 + \frac{2r_-}{r_+}}  \,.
\end{align}
This implies
\begin{equation}
	\sigma = 0\,\quad \xi = -1\,,\quad \kappa^2 = \frac{r_+^3}{r_+ + r_-} \qq{and} \rho = \frac{r_+^2}{r_+ + r_-} \sqrt{1 + \frac{2r_-}{r_+}} \,.
\end{equation}

We are therefore in the case $(\sigma = 0, \xi \leq -1)$ described in Appendix~\ref{App:asymp}, which is the case of the Schwarzschild black hole. One can check that the tortoise coordinate $r_*$ indeed behaves as $\rho \ln(\varepsilon)$ when $\varepsilon$ goes to 0. Furthermore, the behaviour of the potential~\eqref{eq:potential-BCL} as $r_*$ goes to $-\infty$ is indeed given by~\eqref{eq:pot-sigma-0-xi-leq--1}. The arguments of Sec.~\ref{subsec:stability} imply that odd perturbations of the BCL black hole are stable.

\subsubsection{4D Einstein-Gauss-Bonnet solution}

We now turn to another exact black hole, solution of a  DHOST theory obtained  through the compactification of a higher-dimensional Lovelock theory \cite{Lu:2020iav,Hennigar:2020lsl}. The DHOST theory is specified by the following choice of  Lagrangian functions:
\begin{align}
	&F_2 = 1 - 2 \al X \,,\quad P = 2\al X^2 \,,\quad Q = - 4\al X \,,\quad F_3 = -4\al \ln(X) \,,\nonumber\\
	&A_1 = -A_2 = 2 F_{2X} \,,\quad 3 B_1 = -B_2 = \frac32 B_3 = F_{3X} \,,
\end{align}
while all the other  functions are  set to zero. Here, $\al$ is a scalar parameter describing the deviation from GR.

The black hole metric is given by~\eqref{metric} with
\begin{align}
	&\Ab(r) = \Bb(r) = 1 + \frac{r^2}{2\al} \qty(1 - \sqrt{1+\frac{4\al\mu}{r^3}}) = 1 - \frac{\mu}{r} \frac{2}{1+\sqrt{1 + \frac{4\al\mu}{r^3}}} \,,\\
	&\Cb(r)= r^2\,, \qquad \phi(r) = \frac{-1 + \sqrt{\Ab(r)}}{r \sqrt{\Ab(r)}} \,.
\end{align}
The solution reduces to the Schwarzschild black hole in the $\al \longrightarrow 0$ limit: the parameter $\mu$ corresponds to twice the black hole mass in that case.

This black hole has several horizons. The outermost one is located at $r = r_h$, with
\begin{equation}
	r_h = \frac12 \qty(\mu + \sqrt{\mu^2 - 4\al}) \,.
\end{equation}
The perturbations of this black hole solution were studied in \cite{Langlois:2022eta}, and their effective metric was computed in \cite{Langlois:2022ulw}. We can now study the stability of odd parity perturbations of this solution using the framework developped in the present work. 

The study of perturbations of this solution is made easier when one uses the dimensionless quantities given by
\begin{equation}
	z = \frac{r}{r_h} \qq{and} \bet = \frac{\al}{r_h^2} \,.
\end{equation}
With these choices, the parameter $\bet$ is such that $0 \leq \bet \leq 1$ and the outermost horizon is located at $z = 1$. One has therefore
\begin{equation}
	\label{A_z}
	\Ab(z) = 1 + \frac{z^2}{2\bet} \qty(1 - \sqrt{1 + \frac{4\bet(1+\bet)}{z^3}}) = 1 - \frac{2(1+\bet)}{z \left(1+\sqrt{1 + \frac{4\bet(1+\bet)}{z^3}}\right)}\,.
\end{equation}
In the following, we shall work with the quantity $f(z)$ defined from $\Ab(z)$ via
\begin{equation}
	\label{f_z}
	f(z) = \sqrt{\Ab(z)} \,.
\end{equation}
The effective metric obtained in \cite{Langlois:2022ulw} is given by
\begin{equation}
	\Ae = \frac{f^{1/2}}{z^2} \sqrt{\frac{\gamma_1^3 \gamma_2}{\gamma_3^3}} \,,\quad
	\Be = f^{5/2} z^2 \sqrt{\frac{\gamma_3^5}{\gamma_1 \gamma_2^3}} \,,\quad 
	\Ce = f^{-1/2} \sqrt{\frac{\gamma_1 \gamma_2}{\gamma_3}} \,,
\end{equation}
where the functions $\gamma_i$ are defined by
\begin{align}
	\gamma_1 &= f \qty[z^2 + 2\bet(f - 1)(f - 1 - 2 z f') ] \,,\\
	\gamma_2 &= z^4 - 2\bet(1+\bet)z \,,\\
	\gamma_3 &= z^2 + 2\bet(1-f^2) \,.
\end{align}

Choosing $\varepsilon = z - 1$, this implies that the parameters defined in~\eqref{eq:exponents} and~\eqref{asymptotic_parameters} are
\begin{align}
	&\alpha = \frac14 \,,\quad \beta = \frac54 \,,\quad \gamma = -\frac14\,,\quad  \sigma = \frac14\,,\quad \xi = -\frac12 \,, \nonumber\\
	&\rho = (1+2\bet)^{1/4} (1-\bet)^{-5/4} \sqrt{\frac{1}{2\bet} -3\bet -2\bet^2} \,.
\end{align}
We are therefore in the case $\sigma > 0$ and $\xi \leq 2\sigma$. In this situation, the Schr\"odinger operator may or may not be essentially self-adjoint depending on the value of the constant $V_{*s}$ defined in~\eqref{eq:def-Vstars}. For the black hole solution considered here, this constant is\footnote{Since in that case $\nu = -2$, the behaviour of the potential near the horizon is
	\begin{equation}
		V \sim -\frac{1}{4r_*^2} \qq{when} r_* \longrightarrow 0 \,.
\end{equation}}
\begin{equation}
	V_{*s} = \frac{\gamma(\gamma + 2\sigma)}{4 \sigma^2} = -\frac14 < \frac34 \,.
\end{equation}
Henceforth, the Schr\"odinger operator is not essentially self-adjoint which means that we need to choose specific boundary conditions in order to have a uniquely defined time evolution of perturbations. However, as noted in Sec.~\ref{subsec:stability}, this is not necessarily a sign of instability. It may be necessary, in order to compute quasinormal modes of this solution, to impose boundary conditions such that odd perturbations are stable.

\section{Conclusions}
In this work, we have studied the axial perturbations of non-spinning black holes in DHOST theories, the most general family of scalar-tensor theories with a single scalar degree of freedom. Although the family of DHOST theories is huge, the dynamics of axial perturbations can be encapsulated in a few functions that depend on the background metric and on the DHOST functions evaluated on the background. Moreover, there is a  correspondence between the dynamics of axial perturbations in DHOST theories and that of axial perturbations in GR in a different metric, which we call the effective metric. Since we assume that all ordinary fields, like the electromagnetic field, are minimally coupled to the background metric, this implies that axial gravitational waves  propagate (in the ordinary sense) in a different metric compared with other fields.

This effective metric depends explicitly on the choice of the DHOST theory, as well as on the particular background BH solution. Depending on the particular case at hand, the effective metric can either describe another BH geometry, with the same or a different horizon, or a naked singularity. Specific examples of all these possibilities were presented in  \cite{Langlois:2022ulw} 
and are recalled in Section \ref{section:Examples}.

We have studied the effective metric from a general point of view, by assuming a generic power-law behaviour of the effective metric coefficients near the singularity, be it a curvature or a coordinate singularity. Given our ansatz characterized by a few exponents,  we have classified all possible behaviours of the effective tortoise coordinate and of the potential for the effective Schr\"odinger-like equation. This has enabled us to discuss, in a generic way, the self-adjointness status of the Schr\"odinger-like operator and the stability of the perturbations.

Whereas most cases lead to self-adjoint operators, we have nevertheless identified a small region of our parameter space where the operator fails to be self-adjoint. Interestingly, one of our examples turns out to be in this particular region: this is the case of the 4D Einstein-Gauss-Bonnet spacetime. This means that, in this particular case, the time evolution of the axial perturbations seems ambiguously defined, requiring additional information. 

In future work, we would like to extend our analysis to polar perturbations. The situation is however more delicate since the polar perturbations now include an additional mode, that of the scalar field.

\medskip

\begin{acknowledgements}
 K.N. acknowledges support of the National Research Agency (ANR). His research is partially funded by  ANR under the project “ANR-22-CE31-0015-01grant StronG (ANR-22-CE31-0015-01).
 \end{acknowledgements}

\section*{APPENDIX}

\appendix

\section{A new  formulation of the effective metric for axial perturbations}
\label{Section:effective}

In this appendix, we first summarize  some results on the effective metric of axial perturbations obtained in \cite{Langlois:2022ulw}  (see also \cite{Tomikawa:2021pca,Takahashi:2021bml}) and then show how to derive the new (almost covariant) form \eqref{covarianteffective}
for  the effective metric  in  (not necessarily degenerate) higher derivative scalar-tensor theories  including cubic terms.

\subsection{A first expression of the effective metric}

As shown in  \cite{Langlois:2022ulw}, the dynamics  of axial perturbations about any static and spherical symmetric solution of the form \eqref{metric} and \eqref{generalscalar} in any higher order scalar-tensor theory, up to cubic order in second derivatives of the scalar field, can be encapsulated in just four functions ${\cal F}$, $\Phi$, $\Psi$ and $\Gamma$, which depend on the background metric and some of the functions  in the Lagrangian, evaluated on the background. The explicit expressions for these four functions are~\cite{Langlois:2022ulw}
\begin{eqnarray*}
{\cal F} &=&  \Ab \left[ F_2 - X A_1 - Y (F_{3X} + B_2 + XB_6) + (Y -  X \Box \phi) (B_2 + \frac{3}{2}B_3) - \frac{3}{2}X  \phi_t^t B_3 \right]
\\
&  & - q^2 \left[ A_1 + Y B_6 + \Box \phi (B_2 + \frac{3}{2}B_3)  \right] \, ,
\\
\Phi & = &{\cal F}/[ F_2 - X A_1 - Y (F_{3X} + \frac{3}{2} B_3 + X B_6) - \frac{3}{2} \Phi_\theta^\theta B_3 - \frac{3}{2} X \Box \phi B_3] \, ,
\\
	\Psi & = & \frac{q{\cal F}^{-1} }{\Bb \psi'} \left[ X (A_1 + \Box \phi (B_2 + \frac{3}{2}B_3)) + Y (F_{3X} + \frac{3}{2} B_3 + X B_6) \right. \\
		&& \left.  + \frac{q^2}{\Ab} (A_1 +
		\Box \phi (B_2 + \frac{3}{2}B_3) + Y B_6) + \frac{3}{2} B_3 \phi_t^t \right] \, , \\
	\Gamma & = &\Psi^2+ \frac{{\cal F}}{\Ab\Bb} \left[ \Ab F_2 + \Ab Y F_{3X} + q^2 ( A_1 + Y B_6 + \Box \phi (B_2 + \frac{3}{2}B_3) + \frac{3}{2} \Bb \phi_{rr}  B_3 ) - \frac{3}{2}q \Bb \psi' \phi_{rt}\right]\,,
\end{eqnarray*}
where we have introduced the notation
\begin{eqnarray}
\label{Y}
	Y \equiv \phi^\mu  \phi_{\mu\nu} \phi^\nu \, ,
\end{eqnarray}
and used the explicit expressions of the non-trivial second derivatives of the scalar field, namely $\phi_\mu^\nu = g^{\nu \rho}\phi_{\mu\rho}$,
\begin{eqnarray}
	\phi_t^t \, = \, \frac{1}{2} \Bb \frac{\Ab'}{\Ab} \, , \quad
	\phi_r^r \, = \, {\Bb \psi''} - \frac{1}{2} \Bb' \psi' \, , \quad
	\phi_\theta^\theta = \phi_\varphi^\varphi = \frac{1}{2} \Bb \frac{\Cb'}{\Cb} \psi' \, , \quad
	\phi_r^t = \frac{1}{2} q \frac{\Ab'}{\Ab} \, ,
\end{eqnarray}
and the identities
\begin{eqnarray}
	\Box \phi = \phi_t^t + \phi_r^r + 2 \phi_\theta^\theta \, , \quad \phi^\mu X_\mu = \Bb \psi' X' \, , \quad X' = \frac{2}{\Ab} \left( - q \phi_{tr} + \Ab\Bb \psi' \phi_{rr}\right)  \, .
\end{eqnarray}

Remarkably, the DHOST dynamics of the axial perturbations in the background metric is equivalent to the GR dynamics in the effective metric
\begin{eqnarray}
	\label{appmetric}
	\geff_{\mu\nu}\dd x^\mu \dd x^\nu \; = \; \Lambda \left[ - {\cal F} \, \dd t^2 \, + \, N \, \dd r^2 \, + \, 2 P \, \dd r \dd t \, + \, M \, \Cb \, (\dd \theta^2 + \sin^2 \theta \, \dd \varphi^2)\right] \, ,
\end{eqnarray}
where we have introduced 
\begin{eqnarray}
	\label{appGamma}
	\Lambda = \Phi \sqrt{\frac{\Bb}{\Ab} \Gamma} \, , \quad
	N = {\cal F} (\Gamma - \Psi^2) \, , \quad
	P = {\cal F}  \Psi \, , \quad
	M = {\cal F}/\Phi \,.
\end{eqnarray}

From \eqref{appmetric} and \eqref{appGamma}, we immediately deduce the coefficients of the effective metric \eqref{effectivemetricform}
\begin{eqnarray}
	\Ae = -\Lambda {\cal F} \, , \qquad
	\frac{1}{\Be} = N+\frac{P^2}{\cal F} \, , \qquad
	\Ce = \Lambda M \Cb \, ,
\end{eqnarray}
and  the new time coordinate $t_*$ from the relation
\begin{eqnarray}
	\dd t_* = \dd t -   \Psi(r) \, \dd r \, .
\end{eqnarray}

\subsection{A ``more covariant" formulation}
We see that the effective metric involves second derivatives of the scalar field when one considers cubic higher order scalar-tensor theories. This suggests that we can try to  reformulate $\geff_{\mu\nu}$ in the following form,
\begin{eqnarray}
	\geff_{\mu\nu} = \Lambda \left[ \Omega \, \gb_{\mu\nu} \, + \, D \, \phi_\mu \phi_\nu \, + \, S \, \phi_{\mu\nu} + \frac{T}{2} \left( \phi_\mu X_\nu + \phi_\nu X_\mu \right) \right] \, ,
\end{eqnarray}
for some ``covariant'' functions  $\Omega$, $D$, $S$ and $T$ to be determined (in terms of covariant quantities only).
This is possible if these functions satisfy the relations
\begin{eqnarray}
	{\cal F} & = & \Omega \Ab - q^2 D - S \phi_{tt} \, \label{eqforF} \, ,\\
	N & = & \frac{\Omega}{\Bb} + \psi'^2 D + S \phi_{rr} + 2T \psi' \phi^\alpha \phi_{\alpha r} \, ,  \label{eqforN} \\
	P & = & q \psi' D + S \phi_{rt} + q T \phi^\alpha \phi_{\alpha r} \, ,  \label{eqforP}\\
	M & = & \Omega + \frac{S}{C} \phi_{\theta \theta} \, . \label{eqforM}
\end{eqnarray}
Let us underline that these relations impose more than 4 equations for the 4 unknowns as we require that the unknowns take a covariant form as explained above.

The last equation  \eqref{eqforM} together with the expression of $M$ immediately lead to a solution for $\Omega$ and $S$:
\begin{eqnarray}
	\Omega & = & F_2 - X [A_1 + \Box \Phi \, (B_2 + \frac{2}{3} B_3) + Y \, B_6 ]  - Y \,  (F_{3X} - \frac{3}{2} B_3 ) \, , \\
	S & = &  \frac{3}{2} X B_3  \,,
\end{eqnarray}
where we recall that  $Y$ has been defined in (\ref{Y}).
Then, the first equation  \eqref{eqforF} enables us to find the expression of D,
\begin{eqnarray}
	D = A_1 + \Box \Phi \, (B_2 + \frac{2}{3} B_3) + Y \, B_6  \, ,
\end{eqnarray}
while the second one \eqref{eqforN} leads to
\begin{eqnarray}
	T = F_{3X} - \frac{3}{2} B_3 \, .
\end{eqnarray}
Finally, one shows that the third equation \eqref{eqforP} is consistent with the expressions above. As a consequence, we obtain
the desired form of the effective metric with \eqref{expressionofDST}.

\section{Asymptotic behaviour of the potential at the singularity}
\label{App:asymp}
In this appendix, we give more details on the analysis of the asymptotic behaviour of the potential near the singularity.
For that, it is convenient to first decompose the potential into,
\begin{eqnarray}
	V = V_\lambda + V_0 \,  \qquad \text{with} \qquad
	V_\lambda= 2\lambda \frac{\Ae}{\Ce} \, , \qquad V_0= S^2 - \partial_* S \, ,
	\label{eq:def-Vlambda}
\end{eqnarray}
and then  study the behaviours of $V_\lambda$ and $V_0$ separately.

\subsection{Generalities}

As we are going to see, we need to expand the effective metric coefficients, introduced in \eqref{effectivemetricform},  up to sub-leading orders as follows
\begin{eqnarray}
	\label{a1b1c1}
	\Ae(r) \, \simeq \, a \, \varepsilon^\alpha \left( 1+ a_1 \varepsilon \right)\, , \qquad
	\Be(r) \, \simeq \, b \, \varepsilon^\beta \left( 1+ b_1 \varepsilon  \right) \, , \qquad
	\Ce(r) \, \simeq \, c \, \varepsilon^\gamma \left( 1+ c_1 \varepsilon  \right) \, ,
\end{eqnarray}
where ($\alpha$, $\beta$, $\gamma$) and ($a_1$, $b_1$, $c_1$) are real constants, while ($a$, $b$, $c$)  are positive real numbers.
We will also use the notations introduced in \eqref{param2},
\begin{eqnarray}
	\lc^2 \, = \, \frac{c}{a} \, , \qquad \rss \,= \,\frac{\rst}{\sqrt{ab}} \, , \qquad
	\sigma \, = \, 1 - \frac{\alpha+\beta}{2} \, , \qquad
	\xi \, = \, \gamma - \alpha \, .
\end{eqnarray}

The leading order term in the expansion of $V_\lambda$ in powers of $\varepsilon$ is straightforward to obtain, it does not depend on
$(a_1, b_1, c_1)$ and is given by
\begin{eqnarray}
	V_\lambda \simeq 2 \frac{\lambda}{\lc^2} \varepsilon^{-\xi} \, .
\end{eqnarray}
The calculation of the leading order term in the expansion of $V_0$ is subtler.
Indeed, after a direct calculation, one obtains,
\begin{eqnarray}
	V_0 \, \simeq \, V_0^{(0)} \varepsilon^{-2\sigma} + V_0^{(1)}  \varepsilon^{-2\sigma+1} \, ,
\end{eqnarray}
with
\begin{eqnarray}
	V_0^{(0)} \, = \,  \frac{\gamma(\gamma+2\sigma)}{4 \rss^2}  \, , \quad
	V_0^{(1)} = \frac{1}{4 \rss^2} \qty[(a_1 + b_1)(\gamma + 2\sigma - 1) \gamma + 2 c_1 (\gamma + \sigma - 1)] \,.
\end{eqnarray}
We see that $V_0^{(0)}$ vanishes when $\gamma=0$ or $\gamma+2\sigma=0$ and in that case we have to consider the term proportional to $V_0^{(1)}$. Therefore, we will distinguish between the cases  $\gamma \neq 0$ and $\gamma=0$.

In any case, the leading order term in the expansion of the potential in powers of $\varepsilon$ is
\begin{eqnarray}
	\label{Vupsilon}
	V \; \simeq \; V_s \, \varepsilon^\eta \, ,
\end{eqnarray}
where the expressions of $V_s$ and $\eta$ depends on the parameters entering in the metric near the singularity.

\subsection{The case $\gamma \neq 0$}
When $\gamma \neq 0$ (and also $\gamma + 2\sigma \neq 0$), the expressions of $V_s$ and $\eta$ depend on the sign of $\xi-2\sigma$ according to: 
\begin{eqnarray}
	&\text{if} \,\, \xi > 2\sigma \quad ({\rm i.e.}\  \gamma>2-\beta) \, , \qquad  & V_s \, = \, 2\frac{\lambda}{\lc^2}  \, , \quad \eta = -\xi  \, , \\
	&\text{if} \,\, \xi \leq 2 \sigma \quad ({\rm i.e.}\  \gamma\leq 2-\beta)  \, , \qquad & V_s \, = \, \frac{\gamma(\gamma + 2 \sigma)}{4\rss^2} + 2 \frac{\lambda}{\lc^2}  \delta_{\xi-2\sigma}, , \quad \eta =-2\sigma \, ,
\end{eqnarray}
where we introduced the Kronecker symbol $\delta_\xi$ which satisfies $\delta_\xi=0$ if $\xi \neq 0$ and $\delta_0=1$.

In order to express the asymptotic behaviour of the potential in terms of the tortoise coordinate $r_*$, we must distinguish between the
cases $\sigma =0$ and $\sigma \neq 0$.

\medskip

When $\sigma=0$, we find, using \eqref{behaviour_tortoise},
\beq
V \, \simeq \, V_s \, \exp\left( \eta \frac{r_*}{\rss}\right) \,  , \qquad  (\sigma = 0)\,. \label{sigmanul}
\eeq
where $V_s$ and $\eta$ have been given above.
The singularity is located at $r_* \rightarrow - \infty$ and
\begin{eqnarray}
	&&\bullet \;\; \text{if} \, \, \xi >0 \, , \qquad V(r_*) \simeq 2\frac{\lambda}{\lc^2}  \exp \left(-\xi \frac{r_*}{\rss} \right) \, , \\
	&&\bullet \;\; \text{if} \, \, \xi \leq 0 \, , \qquad V(r_*) \simeq  2\frac{\lambda}{\lc^2} \, \delta_\xi + \frac{\gamma^2}{4 \rss^2} \, .
\end{eqnarray}

When $\sigma \neq 0$, we have
\beq
V \, \simeq \, V_{*s} \, \vert r_* \vert^{\nu} \, , \qquad  V_{*s} = V_s \left( \frac{\vert \sigma \vert}{\rss}\right)^{\nu} \, , \qquad \nu =\frac{\eta}{\sigma} \, , 
\qquad (\sigma \neq 0)\,.
\label{sigmanonnul}
\eeq
The singularity is  located at $r_* \rightarrow - \infty$ if $\sigma <0$, at  $r_* =0$ if $\sigma >0$, and the parameters
are given by
\begin{eqnarray}
	&&\bullet \;\; \text{if} \, \, \xi > 2 \sigma \, , \qquad  \nu=-\frac{\xi}{\sigma}  \, , \quad V_s =  2\frac{\lambda}{\lc^2}  \, , \\
	&&\bullet \;\; \text{if} \, \, \xi \leq  2 \sigma \, , \qquad \nu = -2 \, , \quad V_s = 2\frac{\lambda}{\lc^2}  \delta_{\xi - 2\sigma} + \frac{\gamma(\gamma + 2 \sigma)}{4 \rss^2} \, .
\end{eqnarray}

\subsection{The cases $\gamma=0$ and $\gamma+2\sigma=0$}

We proceed as in the previous section but now we compare the two terms 
\beq
	V_\lambda \simeq 2 \frac{\lambda}{\lc^2} \varepsilon^{-\xi}  \quad {\rm and}  \quad V_0 \simeq  V_0^{(1)} \varepsilon^{-2\sigma +1}
\eeq
with
\beq	
	V_0^{(1)}= \frac{(\sigma -1) c_1}{2 \rss^2} \ (\gamma=0)\quad {\rm or} \quad
	V_0^{(1)}= \frac{\sigma (a_1+b_1- c_1)-c_1}{2 \rss^2}  \ (\gamma=-2\sigma)\,.
\eeq
We assume that $\sigma$ is such that the corresponding expression above does not vanish. Otherwise we should consider the next term in the expansion of $V_0$ and the computation would then be similar.

Following the same analysis as before, we still have the behaviour \eqref{Vupsilon} for the potential with different expressions of $V_s$ and $\eta$,
\begin{eqnarray}
	&&\bullet \;\;  \text{if} \,\, \xi > 2\sigma - 1 \,,
	\quad  V_s \, = \, 2\frac{\lambda}{\lc^2}  \, , \quad \eta = -\xi  \, , \\
	&&\bullet \;\;  \text{if} \,\, \xi \leq 2 \sigma-1 \,,
	\quad V_s \, = \,  V_0^{(1)} + 2 \frac{\lambda}{\lc^2} \delta_{\beta + \gamma - 1}\, , \quad \eta =-2\sigma +1 \, .
\end{eqnarray}

We now express the results in terms of $r_*$. The potential takes the same form as in  \eqref{sigmanul} and \eqref{sigmanonnul} with
different parameters.

When $\sigma=0$, the singularity is located at $r_* \rightarrow - \infty$ and
\begin{eqnarray}
	&&\bullet \;\; \text{if} \, \, \xi > -1  \, , \qquad V(r_*) \simeq 2\frac{\lambda}{\lc^2}  \exp \left(-\xi \frac{r_*}{\rss} \right) \, , \\
	&&\bullet \;\; \text{if} \, \, \xi \leq -1 \, , \qquad V(r_*) \simeq  \left( V_0^{(1)} + 2 \frac{\lambda}{\lc^2} \delta_{\xi+1} \right) \exp \left( \frac{r_*}{\rss} \right)\, .
	\label{eq:pot-sigma-0-xi-leq--1}
\end{eqnarray}
Notice that the Schwarzschild metric falls in this case and more precisely it corresponds to $\gamma=0$, $\sigma=0$ (as $\alpha=\beta=1$)
and $\xi=-\alpha=-1$. Therefore, we recover the well-known exponential fall-off of the potential.

When $\sigma \neq 0$, we have,
\begin{eqnarray}
	&&\bullet \;\; \text{if} \, \, \xi > 2 \sigma-1  
	 \, , \qquad  \nu=-\frac{\xi}{\sigma}  \, , \quad V_s =  2\frac{\lambda}{\lc^2}  \, , \\
	&&\bullet \;\; \text{if} \, \, \xi \leq  2 \sigma-1 
	\, , \qquad \nu = -2 + \frac{1}{\sigma}\, , \quad V_s = 2\frac{\lambda}{\lc^2}  \delta_{\beta + \gamma - 1} +V_0^{(1)} \, .
\end{eqnarray}

\medskip

The main differences with the case $\gamma (\gamma + 2 \sigma) \neq 0$ are the existence of new behaviours of the potential which correspond to,
\begin{itemize}
	\item $\sigma=0$ and $-1<\xi<0$ where the potential decreases exponentially when $r_* \rightarrow - \infty$;
	\item $\sigma <0$ where $\nu \geq -2 +1/\sigma$ whereas $\nu \geq -2$ in the previous case, which means the possibility for the potential
	      to tend to zero faster than $1/r_*^2$ when $r_* \rightarrow - \infty$;
	\item $\sigma >0$ where $\nu \leq - 2+1/\sigma$ whereas $\nu \leq -2$ in the previous case, which means  the possibility for the
	      potential $V(r_*) \simeq V_{*s} \, r_*^\nu$ to tend to zero at the singularity.
\end{itemize}

\section{Asymptotic behaviour of the solutions and self-adjointness of the Schr\"odinger operator}
\label{App:SA}
In this section, we summarise some of the results  given in the chapter 2 of  the book \cite{berezin1991one} concerning the
asymptotic behaviour  of the solutions of the Schr\"odinger equation. Then, we apply these general results to the particular cases we are interested in.

\subsection{Asymptotic behaviour of the solution}
Let us consider the one-dimensional Schr\"odinger equation for the wave function $\chi(x)$,
\begin{eqnarray}
	-\frac{\dd^2 \chi}{\dd x^2} + V(x) \chi = \omega^2 \chi \, ,
\end{eqnarray}
where we use the variable $x$ instead of $r_*$ (to lighten the notations).

The authors of the book \cite{berezin1991one}  state a theorem (Theorem 4.6. in Chap. 2) where they give the asymptotic behaviour when
$x \rightarrow + \infty$ of the two solutions of the Schr\"odinger equation depending on the behaviour of the potential itself at infinity.

Let us start with the case where $\vert V(x) \vert \rightarrow \infty$. It is stated \cite{berezin1991one}  that, if the following two integrals
\begin{eqnarray}
	\label{condinfinity}
	\int_{x_0}^\infty \dd x \, \frac{\vert V'(x) \vert^2}{\vert V(x)\vert^{5/2}} \, , \qquad
	\int_{x_0}^\infty \dd x \, \frac{\vert V''(x) \vert}{\vert V(x)\vert^{3/2}}  \, ,
\end{eqnarray}
are convergent for $x_0$ arbitrary large, then there exist a pair $\chi_{\pm}(x)$ of  two independent solutions of the
Schr\"odinger equation whose asymptotic behaviours at infinity are given by,
\begin{eqnarray}\label{equivalenceatinfinity}
	\chi_\pm (x) \simeq \frac{1}{\vert V(x) \vert^{1/4}}  \exp \left[ \pm \epsilon \int \dd x \, \sqrt{\vert V(x) -\omega^2\vert}\right] \, ,
\end{eqnarray}
where $\epsilon=1$ if $V(x) \rightarrow + \infty$ and $\epsilon=i$ if $V(x) \rightarrow - \infty$.

The case where $V(x) \rightarrow 0$ is different. It is stated \cite{berezin1991one}  that, if the following two integrals
\begin{eqnarray}
	\label{conditionat0}
	\int_{x_0}^\infty \dd x \, {\vert V'(x) \vert^2}\, , \qquad
	\int_{x_0}^\infty \dd x \, {\vert V''(x) \vert}  \, ,
\end{eqnarray}
are convergent for $x_0$ arbitrary large,  then there exist a pair $\chi_{\pm}(x)$ of  two independent solutions of the
Schr\"odinger equation whose asymptotic behaviours at infinity are now given by,
\begin{eqnarray}
	\label{equivalenceat0}
	\chi_\pm (x) \simeq  \exp \left[ \pm i \, \omega \int \dd x \, \sqrt{1 -V(x)/\omega^2 }\right] \, .
\end{eqnarray}

\subsection{Application to some potentials}
When, we study the dynamics of axial perturbations, we find Schr\"odinger equivations associated with  potentials which are equivalent (when $x \rightarrow + \infty$) to one of the following,
\begin{eqnarray}
	V(x) = V_0 \exp(\lambda x) \qquad \text{or} \qquad
	V(x) = V_0 \, x^\nu \, .
\end{eqnarray}

\subsubsection{Exponential potentials}
When the potential is exponential, we have two cases to consider. The first one is $\lambda>0$ and $V_0>0$ where $V(x) \rightarrow + \infty$.
We immediately check the conditions \eqref{condinfinity} are satisfied and therefore the two independent solutions $\chi_\pm$ \eqref{equivalenceatinfinity} are  equivalent to
\begin{eqnarray}
	\chi_\pm(x) \simeq  \exp\left[ -\frac{\lambda}{4} x \pm \frac{2 \sqrt{V_0}}{\lambda} \exp \left( \frac{\lambda}{2} x\right)\right] \, .
\end{eqnarray}

The second case is $\lambda<0$ with no conditions on $V_0>0$, hence  $V(x) \rightarrow 0$. The conditions \eqref{conditionat0} are satisfied
and then  the two independent solutions $\chi_\pm$ \eqref{equivalenceat0} are  equivalent to
\begin{eqnarray}
	\chi_\pm(x) \simeq  \exp (\pm i \, \omega x) \, .
\end{eqnarray}

\subsubsection{Power law potentials}
When the potential is a power law, we also encounter two cases. The first one corresponds to $\nu >0$ and $V_0>0$ where  $V(x) \rightarrow + \infty$. The conditions \eqref{condinfinity} are satisfied and therefore the two independent solutions $\chi_\pm$ \eqref{equivalenceatinfinity} are  equivalent to
\begin{eqnarray}
	\chi_\pm(x) \simeq  \, x^{-\nu/4} \, \exp \left[ \pm \frac{2 \sqrt{V_0}}{\nu +2} \, x^{1+\nu/2} \right]  \, .
\end{eqnarray}

The second case corresponds to $\nu <0$ while the sign of $V_0$ is arbitrary, hence $V(x) \rightarrow 0$. The conditions  \eqref{condinfinity} are satisfied and we obtain
\begin{eqnarray}
	\chi_\pm(x) \simeq  \exp (\pm i \, \omega x) \, .
\end{eqnarray}

\bibliographystyle{utphys}
\bibliography{full_biblio}

\end{document}